# Giant superfluorescent bursts from a semiconductor magneto-plasma

G. Timothy Noe II[1], Ji-Hee Kim[1], Jinho Lee[2], Yongrui Wang[3], Aleksander K. Wójcik[3], Stephen A. McGill[4], David H. Reitze[2], Alexey A. Belyanin[3] & Junichiro Kono[1]

[1]Departments of Electrical & Computer Engineering and Physics & Astronomy, Rice University, Houston, Texas 77005, USA, [2]Department of Physics, University of Florida, Gainesville, Florida 32611, USA, [3]Department of Physics and Astronomy, Texas A&M University, College Station, Texas 77843, USA, and [4]National High Magnetic Field Laboratory, Florida State University, Tallahassee, Florida 32310, USA

Currently, considerable resurgent interest exists in superradiance, i.e., a cooperative decay of excited dipoles, originally proposed by Dicke[1].  Recent authors discussed superradiance in diverse contexts – cavity quantum electrodynamics[2], quantum phase transitions[3], and plasmonics[4,5] – where coherent coupling of constituent particles governs their cooperative dynamics.  Here we observe intense, delayed bursts of coherent radiation from a photo-excited semiconductor and interpret it as superfluorescence[6], where macroscopic coherence spontaneously appears from initially incoherent electron-hole pairs.  The coherence then decays superradiantly, with a concomitant abrupt decrease in population from full inversion to zero.  This is the first observation of superfluorescence in a semiconductor, where decoherence is much faster than radiative decay, a situation never encountered in atomic cases.

**Nonetheless, a many-body cooperative state of phased electron-hole "dipoles" does emerge at high magnetic fields and low temperatures, producing giant superfluorescent pulses. The solid-state realization of superfluorescence resulted in unprecedented controllability, promising tunable sources of coherent pulses.**

Quantum particles – electrons, atoms, molecules, electron-hole pairs, and excitons – sometimes cooperate to develop a *macroscopically-ordered state* with extraordinary properties, including superconducting states and Bose-Einstein condensates. In this paper, we discuss an entirely novel setting where a macroscopically-ordered state develops spontaneously: an optically-created dense electron-hole plasma in a semiconductor in a high magnetic field. While optically excited semiconductors have attracted continuing interest for many years through endless discoveries associated with high-density electron-hole pairs, excitons, and polaritons[7-13], the phenomenon we report here has never been observed. Namely, we create an ultradense electron-hole plasma with an intense femtosecond laser pulse, and after a certain delay, an ultrashort burst of coherent radiation emerges, which has a much higher intensity than the spontaneous emission of the same number of electron-hole pairs.

We interpret this striking phenomenon as superfluorescence (SF), i.e., a macroscopic polarization spontaneously develops from an initially incoherent ensemble of electron-hole pairs and abruptly decays producing giant pulses of light. Unlike SF previously observed in atomic systems, all processes occur on picosecond time scales and in a fully controllable fashion as a function of excitation laser power, magnetic field, and temperature, opening up new opportunities for systematic many-body studies and development of novel sources for coherent electromagnetic radiation. We performed theoretical simulations based on the relaxation and recombination dynamics of ultrahigh-density electron-hole pairs in a quantizing magnetic field, which successfully captured the salient features of the experimental observations.

What distinguishes SF from any other light emission processes is the existence of a self-organization stage during which macroscopic coherence, or a giant dipole, builds up spontaneously. This self-phasing stage manifests itself physically as a delay between the excitation pulse and the SF emission pulse. It is purely quantum mechanical in nature, driven by quantum fluctuations, and is a pre-requisite for the subsequent appearance of SF bursts (see Figs. 1a and 1b). Such bursts have enormous intensities, $\propto N^2$ (or $\propto N^{3/2}$ in extended samples[14]), where $N$ is the number of excited dipoles in the ensemble. As Dicke described in his original paper[1], this emission process can be described through the dynamics of a Bloch vector, initially pointing 'north' (all dipoles are in the excited state and there is no coherence) and ending up pointing 'south' (all dipoles are in the ground state and, again, there is no coherence), emitting all its energy as light. The situation is analogous to a classical pendulum at an unstable equilibrium point (see Fig. 1c), for which the rate of change of its angle ($\theta$) from the vertical axis is proportional to $\sin\theta$, initially getting faster with increasing $\theta$, becoming fastest at 90°, then slowing down, and eventually stopping at 180°. This is a totally dissipative process: the initially fully-inverted system is now stable in the ground state after releasing the SF burst of duration much shorter than the decoherence time, which is another distinctive feature of SF.

SF has been observed in atomic gases[15,16] and rarefied impurities in glasses and crystals[17-19], but no direct evidence has been reported for SF using carriers or excitons in solids. Our earlier CW measurements[20,21] suggested the existence of SF in semiconductor quantum wells in high magnetic fields but did not directly detect SF bursts in the time domain. Note that superraddiance studies previously reported for semiconductors[22,23] and polymers[24] are distinctly different from SF – they either deal with an accelerated decay of a single exciton due to translational symmetry breaking in a low-dimensional structure or refer to the cooperative enhancement of recombination rates in *coherently prepared* excitons that do not involve the above-described self-phasing process, an essential ingredient of SF[6,14,25].

The sample used in this study was a stack of fifteen undoped quantum wells consisting of 8-nm $In_{0.2}Ga_{0.8}As$ wells and 15-nm GaAs barriers. The confinement of the well resulted in quantized energy subbands for electrons in the conduction band and holes in the valence band. The strain present in this sample resulted in a large splitting of the heavy-hole ($H_1$) and light-hole ($L_1$) states (see Fig. 2a), with only the heavy-hole states being relevant to the present study. Upon optical excitation using a Ti:Sapphire laser with photon energy centered at 1.55 eV, carriers are excited above the band gap of the GaAs barriers. Both the electrons and holes then experience many scattering events before relaxing into the quantum well to form two-dimensional magneto-excitons. We employ the high-field Landau level (LL) notation, (*NM*), to specify each magneto-exciton state, where $N$ ($M$) is the electron (hole) LL index. The three lowest-energy allowed transitions that we primarily study in this work are the ($NM$) = (00), (11), and (22) transitions, which correspond to the $1s$, $2s$, and $3s$ transitions using the low-field excitonic notation[26].

Pump-probe measurements were made in a transmission geometry in the Faraday configuration (see Fig. 2b), where the pump and probe beams were parallel to the magnetic field and incident normal to the quantum wells. The differential transmission, $\Delta T/T$, when tuned to a particular transition, is proportional to the population inversion for that transition, which is equal to the difference between the number of occupied and unoccupied exciton states. Figures 2c and 2d demonstrate that at the lowest temperature, 5 K, there is a sudden decrease in population inversion when the magnetic field, $B$, is higher than 10 T. At lower $B$, the population dynamics of the (11) transition exhibits a typical long exponential decay, as seen in Fig. 2c. With increasing $B$, the exponential decay transforms into a sudden decrease that becomes faster and occurs at a shorter time delay, ~80 ps, for the (11) transition at 17.5 T. The (22) transition under the same conditions shows similar results except that the sudden decrease in population occurs at an even shorter delay time, ~60 ps, for the highest $B$, as shown in Fig. 2d. Finally, Fig. 2e shows that decreasing the temperature, $T$, and increasing $B$ have a similar effect, i.e., the change in population becomes more sudden and occurs at a shorter time delay when $T$ changes from 150 K to 5 K.

Spectrally and temporally resolved PL, collected in a geometry depicted in Fig. 3a, revealed SF pulses under various $B$, $T$, and pump conditions, as shown in Figs. 3b-3g.

Figure 3b shows TRPL data at 17.5 T and 5 K, spectrally selected for the (00) transition, taken with pump pulse energies, 0.25 nJ and 10 μJ, to be compared with Fig. 1b. For weak excitation (0.25 nJ), the PL shows an initial slow increase due to exciton formation followed by interband relaxation with an exponential decay time of hundreds of ps. The PL measured from the edge fiber provided a similar decay, but the signal was ~40 times lower, indicating that the emission under weak excitation is typical spontaneous emission radiated in all directions with equal probability. In contrast, for strong excitation (10 μJ), we observe a giant, delayed pulse of radiation from the edge fiber. The PL measured in the center fiber showed no pulse of radiation, and the peak intensity was ~100 times lower. Quantitatively, Fig. 3b shows that an increase in pump pulse energy by roughly 4 orders of magnitude results in a peak emission intensity that is roughly 6 orders of magnitude larger. This is in agreement with the expected $N^{3/2}$-dependence for SF in extended samples[14], where $N$ is the number of excited dipoles (electron-hole pairs in the present case). These results indicate that with increasing magnetic field strength, increasing pump pulse energy, and decreasing temperature the regime of light emission transitions from ordinary spontaneous emission to SF dominating the in-plane emission at high pump-pulse energy. Figure 3c overlays pump-probe and TRPL data taken under the same conditions for the (22) transition, where we see that the appearance of the giant emission pulse coincides in time with the abrupt population drop from its maximum value to zero. This is in stark contrast with the dynamics of ordinary single-pass amplifiers, where a pulse of the amplified spontaneous emission would consume at most half of the population.

Figure 3d shows a PL intensity map as a function of delay time and photon energy at 17.5 T, 5 K, and 5 μJ.   SF pulses coming from the (00), (11), and (22) transitions are clearly resolved both in time and energy.   For each transition, a large pulse of radiation appears after some delay time.   The highest-energy transition, (22), emits a pulse first, and each lower-energy transition emits a pulse directly after the transition just above it.   Figure 3e shows the effects of lowering $B$ to 15 T: i) the separation between LLs decreases (compare the left hand panels of Figs. 3d and 3e), ii) the SF emission occurs at later delay times for a given transition (compare the right hand panels of Figs. 3d and 3e), and iii) the SF pulse intensity decreases (compare Figs. 3d and 3e).   With decreasing pump pulse energy, we see the emission decrease dramatically (compare Figs. 3d and 3f).   With increasing $T$, the emission from all transitions weakens significantly and moves to later delay times (see Fig. 3g); the emission energies decrease with increasing $T$ due to band gap shrinkage.

These pump-probe and TRPL results are consistent with the expected conditions required to make SF emission possible.   In particular, the $B$ and $T$ dependence of the population drop and SF emission intensity can be explained as follows.   The key parameter determining the growth rate of SF is the cooperative frequency, $\omega_c$, which must exceed the dephasing rate of the optical polarization associated with a given recombination transition for SF to be observable[14].   Specifically, to observe SF from electron-hole pairs in a semiconductor quantum well,

$$\omega_c = \sqrt{\frac{8\pi^2 d^2 n \Gamma c}{\hbar \tilde{n}^2 \lambda L_{QW}}} \geq \frac{2}{T_2} \qquad \dots (1)$$

must be satisfied[20,21], where $d$ is the transition dipole moment, $n$ is the 2D electron-hole density, $\Gamma$ is the overlap factor of radiation with the quantum wells, $\hbar$ is the reduced Planck constant, $\tilde{n}$ is the refractive index, $\lambda$ is the wavelength, $c$ is the speed of light, $L_{QW}$ is the total width of the quantum wells, and $T_2$ is the dephasing time of the optical polarization. A magnetic field increases $d$, $n$, and $T_2$, making it easier to establish the macroscopic coherence required for SF emission. Quantitatively, $d$ increases by a factor of 3 for the (22) transition when $B$ increases from 0 T to 17 T, due to an increased overlap of electron and hole wave functions, and thus, the growth rate in the active region increases to 5-10 ps$^{-1}$. Increasing $T$ decreases $T_2$, making it more difficult to establish the macroscopic coherence. A high $B$ and low $T$ strongly suppress all scattering processes in the lowest LLs because these states are nearly completely occupied with carriers whereas empty states are located high above. Under these conditions one can expect $T_2$ to be on the scale of tens of ps.

We simulated the dynamics of SF in quantum wells by calculating the eigenstates and eigenfunctions of the relevant states, deriving the matrix elements of the optical transitions and phonon scattering, and solving the coupled density-matrix and wave equations for exciton populations, interband coherences, and optical fields (see Methods Summary for details). Figure 4 shows the dynamics of electron-hole pair population and edge-propagating radiation intensity calculated for the (22) state at 17 T. Here, we assumed a $T_2$ of 50 ps (or a dephasing rate of 0.02 ps$^{-1}$), while the calculated modal growth rate of SF for this case was 0.3 ps$^{-1}$.   As seen in Fig. 4, all states are fully occupied within the first few ps with their population $n$ saturating, corresponding to the plateaus observed in the pump-probe measurements of Fig. 2.   After a delay time of about 60 ps, a SF pulse emerges, with a duration of 10-20 ps (shorter than $T_2$).   The emission consumes nearly all the population, bringing the effective population inversion of the (22) level $\Delta n = 2n - 1$ from +1 to -1.   Note that a pulse of amplified spontaneous emission would consume only half of the population, bringing $n$ to ½ and $\Delta n$ to zero.   Overall, there is excellent agreement between the simulations and the pump-probe and time-resolved PL data of Fig. 3. The dynamics of SF from (11) and (00) Landau levels is similar to that of the (22) level, except that the SF pulse from lower Landau levels develops at later times and has a higher intensity, again in agreement with the experiment.

It is important to point out that, while there is a certain analogy between recombination of electron-hole pairs (or excitons) and radiative transitions in atoms, there is no *a priori* reason for SF in semiconductors to be similar to atomic SF (or even to exist at all).   In fact, all previous SF observations were made in low-density atomic systems, where any decoherence processes were negligible and the only relevant decay was via radiative coupling.   In the present case, however, we deal with an ultradense electron-hole plasma in a semiconductor, a complex many-body system with a variety of ultrafast interactions.   The decoherence rates are at least 1,000 times faster than the rate of the radiative decay of individual excitons, a new regime totally unexplored in previous atomic SF studies.   We have shown, nonetheless, that collective many-body coupling of electron-hole pairs via a common radiation field does develop under certain conditions and leads to the spontaneous formation of a macroscopic optical polarization from an initially completely incoherent state.

Finally, the solid-state realization of SF in the current work brought an unprecedented degree of controllability in the generation of SF, opening up new opportunities for both fundamental many-body studies and device applications. We clearly demonstrated that the intensity, duration, and delay time of SF bursts are tunable through $B$, $T$, and pump laser power. Namely, unlike atoms, we can tune virtually everything: the density of states, the oscillator strength, the transition frequencies, and, most importantly, the decoherence times via $B$ and $T$. The fact that $B$ affects both the characteristics of SF and the strength of Coulomb interaction can lead to interesting aspects of SF in a way that other systems cannot. Furthermore, we can make use of advanced semiconductor technology and design compact SF-based devices, contingent on the degree to which the strong $B$ and low $T$ requirements can be lessened by sample design improvements. For example, by increasing the number of quantum wells or surrounding them with symmetric cladding layers, the electromagnetic SF modes can be guided in the active region, increasing the modal overlap and enhancing the SF growth rate to the order of 3-5 ps$^{-1}$. In this case, one should expect to observe much shorter pulses and delay times, and SF should survive at higher dephasing rates and become observable even at room temperature. Eventually, an electrically-driven device for producing coherent SF pulses with any desired wavelength could be developed by utilizing existing technologies of semiconductor quantum engineering.

**Methods**

**Pump-probe and time-resolved PL measurements.** We used an amplified Ti:Sapphire laser as the pump and a tunable optical parametric amplifier (OPA) to probe the population inversion of the (00), (11), and (22) transitions as a function of time using standard delay stage pump-probe techniques. The sample was placed in a 17.5 T superconducting magnet in the Faraday geometry where the field was parallel to the optical excitation and perpendicular to the quantum well plane. The pump and probe beams were made collinear before entering the magnet bore through a $CaF_2$ window. The probe was collected with an optical fiber to bring the light out of the magnet and then filtered using a small monochromator before being collected with a silicon photodiode and measured with a lock-in amplifier at the modulation frequency of an optical chopper introduced in the path of the pump line. We measured the time-resolved PL using a streak camera with 2-ps resolution. The sample was attached to a sapphire window, and a right angle micro prism was mounted at the edge of the sample to redirect the in-plane emission into an edge collection fiber (see Fig. 3a). The output of the fibers was f-number matched with a spectrometer so that the PL was spectrally separated before entering the streak camera.

**Modeling.** Theoretical modeling of SF starts from calculating eigenstates and eigenfunctions for electron-heavy hole quantum-well magneto-excitons following the steps outlined in Ref. 27. These eigenfunctions are used to derive matrix elements of the optical transitions and phonon scattering rates. Then we solve a coupled set of space- and time-dependent density-matrix and wave equations for exciton populations, interband coherences (off-diagonal density-matrix elements), and optical fields. Since all the photo-created carriers eventually occupy the lowest Landau levels to near-complete degeneracy, before the SF pulses are formed, we include only the (00), (11), and (22) states. The initial kinetics of carrier cooling and relaxation are modeled by an effective scattering rate to the (22) state, which is assumed to be a pulse of amplitude 0.1-1 ps$^{-1}$ with an exponential decay time of ~10 ps. The dominant interaction between excitons that leads to the formation of SF pulses is via their coupling to a common electromagnetic mode. Other interactions between excitons are ignored since they are not essential to the SF dynamics, although they renormalize transition energies and Rabi frequencies.

**Full Methods** are available in the online version of the paper at www.nature.com/naturephysics.

**Acknowledgements**

This work was supported by the National Science Foundation through Grants DMR-1006663 and ECS-0547019. A portion of this work was performed at the National High Magnetic Field Laboratory, supported by NSF Co-operative Agreement No. DMR-0084173 and by the State of Florida. We thank Glenn Solomon for providing us with the InGaAs/GaAs quantum well sample used in this study.

**Author contributions**

G.T.N., J.H.K., and J.L. performed the measurements presented in this manuscript, in collaboration with S.A.M. J.L. did most of the initial work of setting up the streak camera. Y.W., A.W., and A.A.B. developed the theoretical model and performed simulations. D.H.R. and J.K. provided overall supervision and guidance on the experimental aspects. All authors contributed to data analysis and interpretation as well as the writing of the manuscript.

**Additional information**

The authors declare no competing financial interests. Supplementary information accompanies this paper in [www.nature.com/naturephysics](www.nature.com/naturephysics). Reprints and permissions information is available online at [www.nature.com/reprints](www.nature.com/reprints). Correspondence and requests for materials should be addressed to J.K. ([kono@rice.edu](mailto:kono@rice.edu)).

# Figure Legends

**Figure 1 | Superfluorescence (SF) from a collection of dipoles (atoms, molecules, ions, or excitons). a,** Self-organization of dipoles and resulting SF pulse. **b,** Characteristics of light emission dynamics after pulse excitation for weak (left) and strong (right) excitation. When the number of dipoles, $N$, is smaller than a critical value ($N_c$), the peak intensity is proportional to $N$ and the intensity decays exponentially with a lifetime $T_1$.   Under high excitation such that $N > N_c$, a delayed SF pulse appears with intensity proportional to $N^2$ and pulse width ~ $T_1/N$. **c,** Population inversion and emitted light intensity versus time for an SF system, together with Bloch vector dynamics analogous to the dynamics of an over-damped pendulum going from an unstable equilibrium position ($\theta = 90°$) to the stable ground state ($\theta = 0°$) by releasing all energy as an SF burst.

**Figure 2 | Observation of a sudden population drop through ultrafast pump-probe spectroscopy. a,** Sample studied and schematic diagram of energy levels in the system. **b,** Experimental configuration of pump-probe measurements. **c,** Pump-probe data for (11) level at different magnetic fields at 5 K. **d,** Pump-probe data for (22) level at different magnetic fields at 5 K. **e,** Pump-probe data for (22) level at 17.5 T at different temperatures.

**Figure 3 | Observation of delayed bursts of radiation through time-resolved photoluminescence spectroscopy. a,** Setup. **b,** Time-resolved emission for weak (top) and

strong (bottom) excitation for (00) transition at 17.5 T and 5 K.   This should be compared with Fig. 1b. **c,** Comparison between pump-probe data and time resolved photoluminescence for (22) level at 17.5 T and 5 K, demonstrating the temporal coincidence between the population drop in pump-probe differential transmission and emission of the giant pulse of radiation.   This should be compared with Fig. 1c.   Streak camera images of emission intensity as a function of photon energy and delay time at **d,** 17.5 T, 5 μJ pump pulse energy, and 5 K, **e,** 15.0 T, 5 μJ, and 5 K, **f,** 17.5 T, 2 μJ, and 5 K, and **g,** 17.5 T, 5 μJ, and 75 K.   The left hand panels of **d-g** show the time-integrated emission spectra with the (00) peak in red, (11) in blue, and (22) in black, while the right hand panels show time-resolved slices at the peak positions of the (00), (11), and (22) transitions.

**Figure 4 | Theoretical simulations of superfluorescence from an ultradense electron-hole plasma in a semiconductor quantum well in a perpendicular magnetic field of 17 T.** Occupation number of excitons on the (22) level (blue dashed line) and normalized electromagnetic field intensity of superfluorescence from the (22) level (red solid line) as a function of time since the beginning of the pump pulse.   Population equal to 1 corresponds to total inversion where all states are fully occupied by excitons.

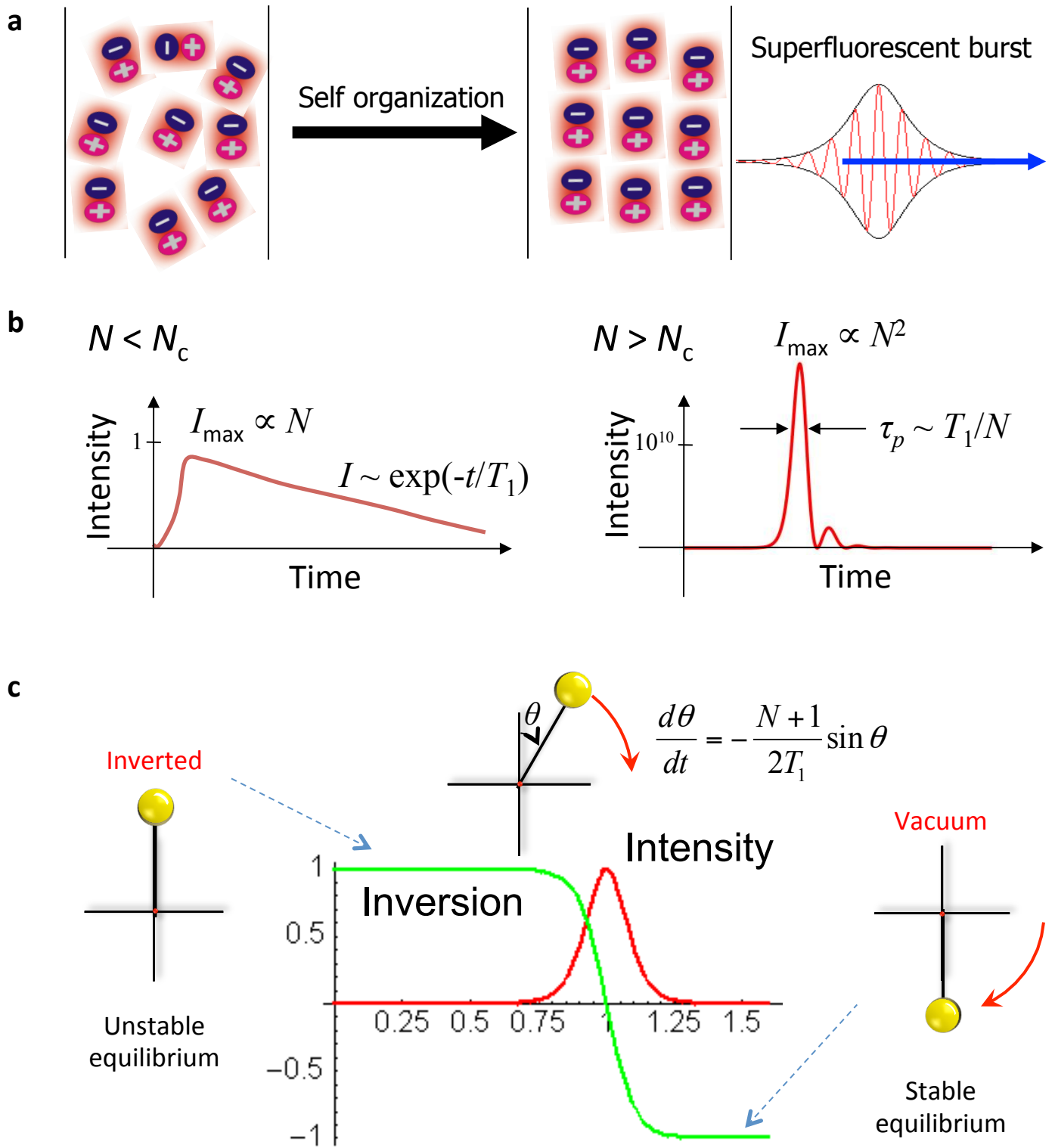

FIG. 1

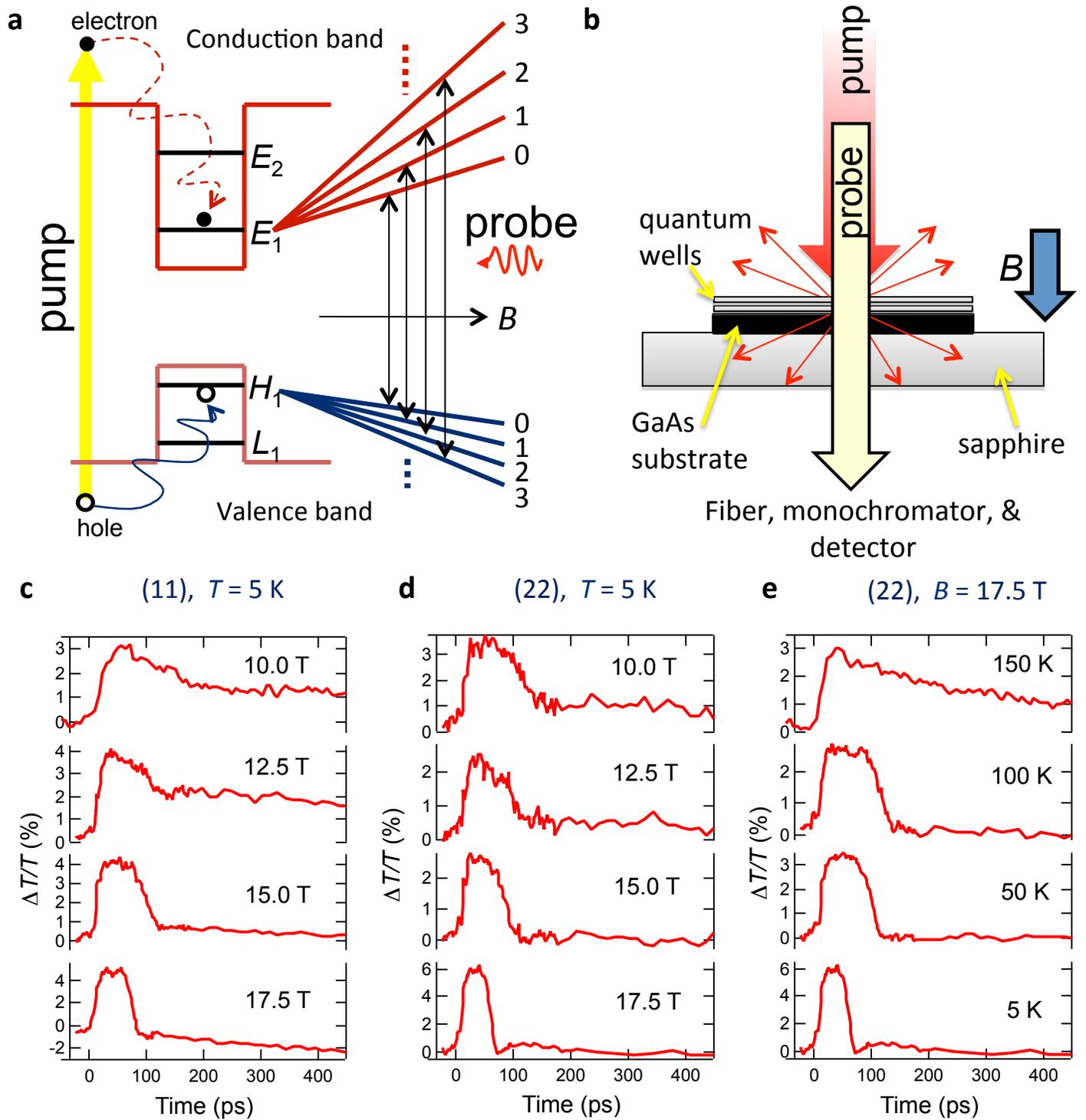

FIG. 2

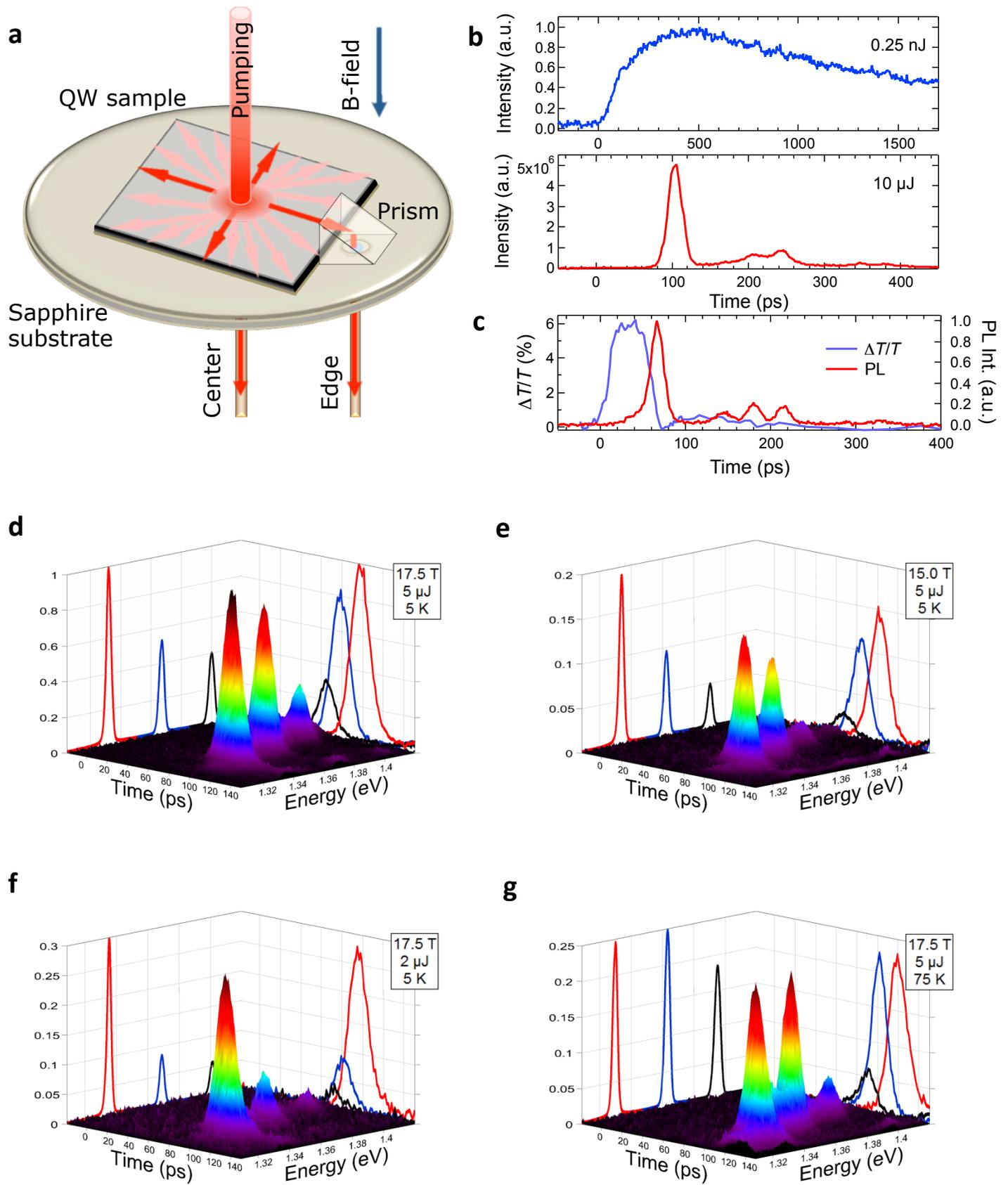

FIG. 3

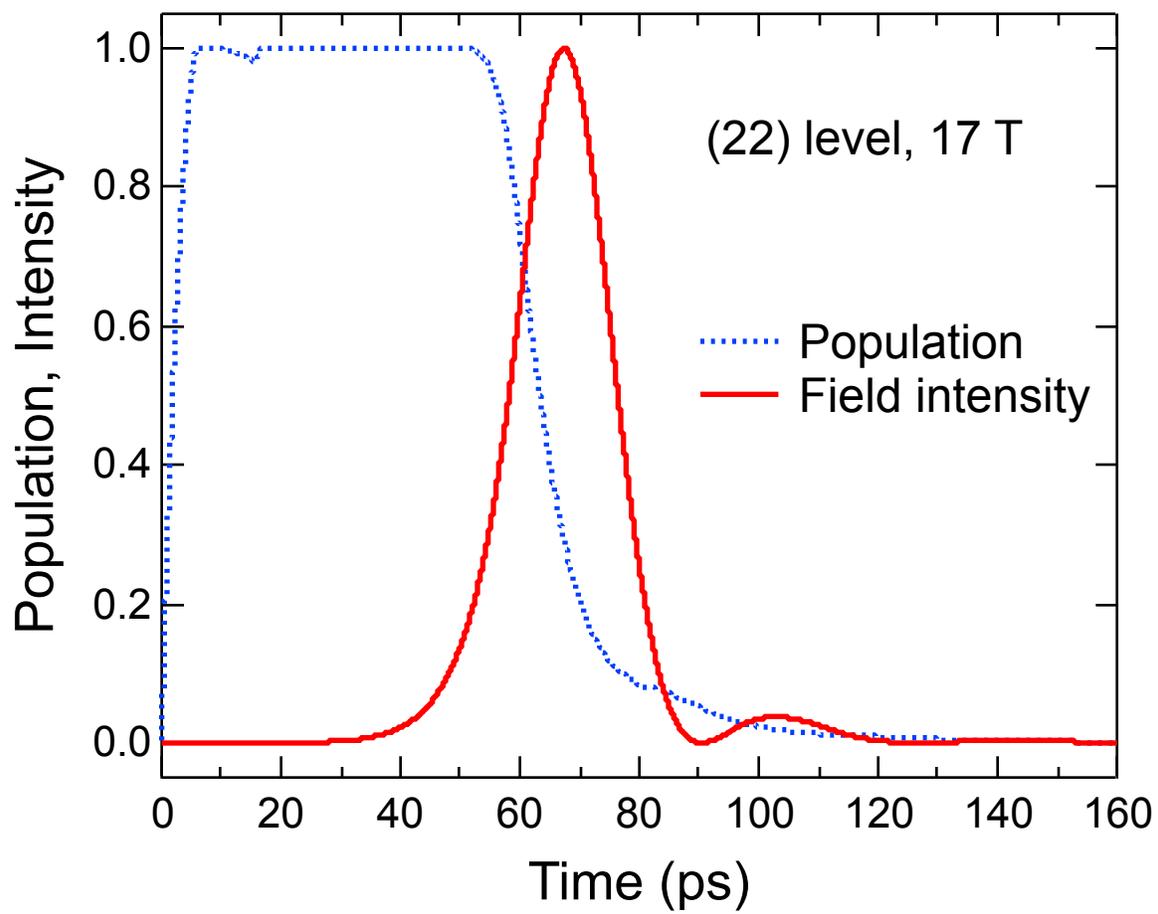

FIG. 4